\documentstyle[12pt]{article}
\catcode`\@=11
\ifcase \@ptsize
\or % mods for 11 pt
\or % mods for 12 pt
\fi
\advance\textheight by \topskip
\ifcase \@ptsize
\or % mods for 11 pt
\or % mods for 12 pt
\fi
\begin{document}
\flushbottom
\pagestyle{empty}
\setcounter{page}{0}
\begin{flushright}
DFTT 1/95
\\
hep-ph/9501241
\end{flushright}
\vspace*{1cm}
\begin{center}
\Large \bf
A Model Independent Approach
\\
\vspace{0.3cm}
to Future Solar Neutrino
Experiments$^{\mbox{\footnotesize\mediumseries$\star$}}$
\vspace*{1cm}
\\
\large \mediumseries
S.M. Bilenky$^{\mbox{\footnotesize\mediumseries(a,b)$\dagger$}}$
and
C. Giunti$^{\mbox{\footnotesize\mediumseries(a)$\ddagger$}}$
\\
\vspace{0.5cm}
\large
$^{\mbox{\footnotesize\mediumseries(a)}}$
INFN, Sezione di Torino,
and
Dipartimento di Fisica Teorica,
\\
Universit\`a di Torino,
Via P. Giuria 1, I--10125 Torino, Italy
\\
$^{\mbox{\footnotesize\mediumseries(b)}}$
Joint Institute for Nuclear Research, Dubna, Russia
\\
\vspace*{1cm}
Abstract
\\
\vspace{0.5cm}
\normalsize
\begin{minipage}[t]{0.9\textwidth}
It is shown that
from the data of future solar neutrino experiments
(SNO and Super-Kamiokande),
in which high-energy
$^8\mathrm{B}$ neutrinos will be detected,
it will be possible
in a model independent way:
1) To reveal the presence of sterile neutrinos
in the flux of solar neutrinos on the earth;
2) To obtain the initial flux of
$^8\mathrm{B}$ $\nu_e$'s;
3) To determine the probability
of $\nu_e$'s to survive
as a function of neutrino energy.
\end{minipage}
\end{center}
\vfill
\vspace*{1cm}
\begin{flushleft} \footnotesize
$\mbox{}^{\mbox{\footnotesize\mediumseries$\star$}}$
Talk presented by S.M. Bilenky at the Conference
{\it Beyond the Standard Model IV},
Lake Tahoe, California,
December 13-18, 1994.
\\
$\mbox{}^{\mbox{\footnotesize\mediumseries$\dagger$}}$
E-mail address: BILENKY@TO.INFN.IT
\\
$\mbox{}^{\mbox{\footnotesize\mediumseries$\ddagger$}}$
E-mail address: GIUNTI@TO.INFN.IT
\end{flushleft}

\newpage
\pagestyle{plain}

Solar neutrino experiments
are a powerful tool
for the investigation of
the problem of neutrino mixing
and
for the search of new physics beyond the Standard Model.
It is well known that
from the analysis of the existing data
follow some indications in favor
of non-zero neutrino masses and mixing~\cite{B:GALLEX92B}.
This analysis is based,
however,
on the assumption that the
Standard Solar Model~\cite{B:SSM} (SSM)
correctly predicts
the solar neutrino fluxes
from the different reactions of the
solar thermonuclear cycles.

In this report
we will consider the
SNO~\cite{B:SNO}
and
Super-Kamiokande~\cite{B:SK} (S-K)
experiments,
scheduled to start in 1995-96.
We will show that these experiments
will allow
to investigate neutrino mixing
in a model independent way
and
to determine the initial $^8\mathrm{B}$ neutrino flux
directly from the data.

In the SNO experiment
solar neutrinos will be detected through the observation
of the CC, NC
and
CC+NC elastic scattering (ES)
processes:
\begin{eqnarray}
&&
\nu_{e} + d \to e^{-} + p + p
\;,
\label{E:CC}
\\
&&
\nu + d \to \nu + p + n
\;,
\label{E:NC}
\\
&&
\nu + e^{-} \to \nu + e^{-}
\;.
\label{E:ES}
\end{eqnarray}
In the S-K experiment
solar neutrinos will be detected through
the observation of the process (\ref{E:ES}),
with an event rate that is expected to be about 50 times
larger than in the current Kamiokande experiment.
Due to the high energy thresholds,
only high energy $^8\mathrm{B}$ neutrinos
will be detected in both experiments.
The initial flux of $^8\mathrm{B}$ $\nu_{e}$'s
is given by
\begin{equation}
\phi_{\nu_{e}}^{0}(E)
=
X(E)
\,
\Phi
\;,
\label{E500}
\end{equation}
where $X(E)$ is a known function
(determined by the phase space factor of the decay
$ ^8\mathrm{B} \to \mbox{} ^8\mathrm{Be} + e^{+} + \nu_{e} $)
and
$\Phi$
is the total flux.

As it is well known
(see, for example, Ref.\cite{B:BILENKY}),
a discovery of transitions of active neutrinos
into sterile states
would be a discovery of new physics
beyond the Standard Model.
We will show first
that the SNO and S-K experiments
could allow to reveal
the presence of such transitions
in a model independent way.

Let us consider the NC process (\ref{E:NC}).
For the average total
probability
of transitions of $\nu_{e}$
into all active neutrinos we have
\begin{equation}
\left\langle
\sum_{\ell=e,\mu,\tau} \mathrm{P}_{\nu_{e}\to\nu_{\ell}}
\right\rangle_{\mathrm{NC}}
=
{\displaystyle
N^{\mathrm{NC}}
\over\displaystyle
X_{{\nu}d}
\,
\Phi
}
\;,
\label{E503}
\end{equation}
where
$ N^{\mathrm{NC}} $
is the NC event rate
and
\begin{equation}
X_{{\nu}d}
\equiv
\int_{E_{\mathrm{th}}^{\mathrm{NC}}}
\sigma_{{\nu}d}(E) \,
X(E) \,
{\mathrm{d}} E
\simeq
4.72 \times 10^{-43} \,\mathrm{cm}^2
\;.
\label{E599}
\end{equation}
In the general case of transitions
of $\nu_e$'s
into active as well as into sterile neutrinos
\begin{equation}
\sum_{\ell=e,\mu,\tau} \mathrm{P}_{\nu_{e}\to\nu_{\ell}}(E)
=
1
-
\mathrm{P}_{\nu_{e}\to\nu_{\mathrm{s}}}(E)
\;,
\label{E504}
\end{equation}
where
$ \mathrm{P}_{\nu_{e}\to\nu_{\mathrm{s}}}(E) $
is the total
probability of transition of
$\nu_e$'s
into all possible sterile states.
It is clear from Eqs.(\ref{E503}) and (\ref{E504})
that
from the measurement of
$ N^{\mathrm{NC}} $
it is impossible to reach any conclusions
about $\nu_e\to\nu_{\mathrm{s}}$
transitions
without an assumption
on the value of the total flux
$ \Phi $.

Let us take into account,
however,
that in the future SNO and S-K experiments
solar neutrinos
will be detected also through the observation of
CC and CC+NC reactions.
We have
\begin{equation}
\left\langle
\sum_{\ell=e,\mu,\tau} \mathrm{P}_{\nu_{e}\to\nu_{\ell}}
\right\rangle_{\mathrm{ES}}
=
{\displaystyle
\Sigma^{\mathrm{ES}}
\over\displaystyle
X_{\nu_{\mu}e}
\,
\Phi
}
\;.
\label{E509}
\end{equation}
Here
\begin{equation}
X_{\nu_{\mu}e}
\equiv
\int_{E_{\mathrm{th}}^{\mathrm{ES}}}
\sigma_{\nu_{\mu}e}(E)
\,
X(E)
\simeq
2.08 \times 10^{-45} \, \mathrm{cm}^2
\,
{\mathrm{d}} E
\label{E510}
\end{equation}
and
\begin{equation}
\Sigma^{\mathrm{ES}}
\equiv
N^{\mathrm{ES}}
-
\int_{{E_{\mathrm{th}}^{\mathrm{ES}}}}
\left(
\sigma_{\nu_{e}e}(E)
-
\sigma_{\nu_{\mu}e}(E)
\right)
\phi_{\nu_{e}}(E)
\,
{\mathrm{d}} E
\;,
\label{E508}
\end{equation}
where
$ N^{\mathrm{ES}} $
is the ES event rate
and
$ \phi_{\nu_{e}}(E) $
is the flux of $\nu_e$ on the earth,
which will be determined
in the SNO experiment
from the measurement
of the electron spectrum
in the CC reaction (\ref{E:CC}).
{}From Eqs.(\ref{E503}) and (\ref{E509})
we obtain the following lower bounds:
\begin{eqnarray}
&&
\left\langle
\mathrm{P}_{\nu_{e}\to\nu_{\mathrm{s}}}
\right\rangle_{\mathrm{ES}}
\ge
1
-
\mathrm{R}^{\mathrm{ES}}_{\mathrm{NC}}
\;,
\label{E516}
\\
&&
\left\langle
\mathrm{P}_{\nu_{e}\to\nu_{\mathrm{s}}}
\right\rangle_{\mathrm{NC}}
\ge
1
-
\left( \mathrm{R}^{\mathrm{ES}}_{\mathrm{NC}} \right)^{-1}
\;,
\label{E517}
\end{eqnarray}
where
\begin{equation}
\mathrm{R}^{\mathrm{ES}}_{\mathrm{NC}}
\equiv
{\displaystyle
\Sigma^{\mathrm{ES}}
\,
X_{{\nu}d}
\over\displaystyle
X_{\nu_{\mu}e}
\,
N^{\mathrm{NC}}
}
\; .
\label{E513}
\end{equation}
is a
{\em measurable}
quantity
(that does not depend on $\Phi$).
Thus,
if
$ \mathrm{R}^{\mathrm{ES}}_{\mathrm{NC}} \not= 1 $
it will mean that sterile neutrinos exist
and Eq.(\ref{E516}) or Eq.(\ref{E517})
will give
a lower bound for the average
$ \nu_{e}\to\nu_{\mathrm{s}} $
transition probability.
Let us notice that
in the case
$ \mathrm{R}^{\mathrm{ES}}_{\mathrm{NC}} = 1 $
no conclusion about sterile neutrinos
can be reached.

Several relations analogous to
(\ref{E516}) and (\ref{E517}),
which will allow to test
in a model independent way
whether
there are transitions of solar $\nu_e$'s
into sterile states,
were derived in Ref.\cite{B:ARTIC}.
If all of them will not reveal
the presence of
$ \nu_{e}\to\nu_{\mathrm{s}} $
transitions,
it will be natural to assume that
$ \displaystyle
\sum_{\ell=e,\mu,\tau} \mathrm{P}_{\nu_{e}\to\nu_{\ell}}(E)
=
1
$.
In this case
the total $^8\mathrm{B}$ neutrino flux
can be determined directly
from the experimental data.
{}From Eqs.(\ref{E503}) and (\ref{E509})
we obtain
\begin{equation}
\Phi
=
{\displaystyle
N^{\mathrm{NC}}
\over\displaystyle
X_{{\nu}d}
}
\qquad
\mbox{and}
\qquad
\Phi
=
{\displaystyle
\Sigma^{\mathrm{ES}}
\over\displaystyle
X_{\nu_{\mu}e}
}
\;.
\label{E567}
\end{equation}
A comparison
of the flux $ \Phi $
obtained from Eq.(\ref{E567})
with the $^8\mathrm{B}$ neutrino flux
predicted by the SSM
will be an important test of the model.

The SNO and S-K experiments
will allow
to determine the $\nu_e$
survival probability
directly from measurable quantities:
\begin{equation}
\begin{array}{l} \displaystyle
\mathrm{P}_{\nu_{e}\to\nu_{e}}(E)
=
{\displaystyle
\phi_{\nu_{e}}(E)
\over\displaystyle
X(E)
\,
\Phi
}
\;,
\\ \displaystyle
\left\langle
\mathrm{P}_{\nu_{e}\to\nu_{e}}
\right\rangle_{\mathrm{ES}}
=
{\displaystyle
1
\over\displaystyle
X_{\nu_{e}e}
-
X_{\nu_{\mu}e}
}
\left[
{\displaystyle
N^{\mathrm{ES}}
\over\displaystyle
\Phi
}
-
X_{\nu_{\mu}e}
\right]
\;.
\end{array}
\label{E570}
\end{equation}
Here
$ \phi_{\nu_{e}}(E) $
is the flux of solar $\nu_e$'s
on the earth
that will be measured by SNO
and
the total flux
$ \Phi $
is given by
Eq.(\ref{E567}).

In conclusion,
we have shown that
future solar neutrino experiments
will allow
to obtain model independent informations
about neutrino mixing
and
the initial solar $\nu_e$ flux.
These experiments
have a good potential
for revealing new physics
beyond the Standard Model.

\end{document}